\documentclass[amsmath, amsfonts, superscriptaddress, showpacs, twocolumn, prl]{revtex4}
\usepackage{graphicx}
\usepackage{epsfig}
\usepackage{bm}
\usepackage{dcolumn}
\usepackage{amsmath}
\usepackage{amssymb}
\usepackage{euscript}

\begin{document}

\title{Thermalization of nonequilibrium electrons in quantum wires}

\author{Tobias Micklitz}
\affiliation{Dahlem Center for Complex Quantum Systems and Institut
f\"{u}r Theoretische Physik, Freie Universit\"{a}t Berlin, 14195
Berlin, Germany}

\author{Alex Levchenko}
\affiliation{Materials Science Division, Argonne National
Laboratory, Argonne, IL 60439, USA} \affiliation{Department of
Physics and Astronomy, Michigan State University, East Lansing, MI
48824, USA}

\begin{abstract}
We study the problem of energy relaxation in a one-dimensional
electron system. The leading thermalization mechanism is due to
three-particle collisions. We show that for the case of spinless
electrons in a single channel quantum wire the corresponding
collision integral can be transformed into an exactly solvable
problem. The latter is known as the Schr\"odinger equation for a
quantum particle moving in a P\"oschl-Teller potential. The spectrum
for the resulting eigenvalue problem allows for bound-state
solutions, which can be identified with the zero modes of the
collision integral, and a continuum of propagating modes, which are
separated by a gap from the bound states. The inverse gap gives the
time scale at which counterpropagating electrons thermalize.
\end{abstract}

\date{September 20, 2011}


\maketitle

{\it Introduction.---}Relaxation processes and nonequilibrium
dynamics in one-dimensional electron systems have moved into the
focus of recent
theoretical~\cite{gutman,bagrets,takei,torsten,Lunde,JK,JTKA,QHE-equilibration,Heyl}
and experimental research~\cite{chen,granger,barak,altimiras}. One
essential characteristic of one-dimensional electrons is the absence
of relaxation in the case of a linear energy dispersion relation and
a slow relaxation if dispersion is nonlinear, inhibited by momentum
and energy-conservation laws~\cite{torsten,JTKA}. Based on
momentum-resolved tunneling spectroscopy the peculiarities of
relaxation in one-dimesional electron systems have been observed in
recent experiments~\cite{chen,barak}. Moreover, the violation of the
Wiedeman-Franz law observed~\cite{chiatti} at the plateau of the
electrical conductance in single channel quantum wires can be
qualitatively understood from the different relaxation processes
required for equilibration of applied temperature gradients and
chemical potentials.

In the presence of many transport channels relaxation of electrons
at low temperature is primarily provided by pair collisions. A
solution of the eigenvalue equation of the two-particle collision
integral in two and three dimensions has already been given four
decades ago at the early era of Fermi liquid theory~\cite{KEFL}. A
remarkable result of that theory is that the eigenvalue equation
turns out to be exactly solvable. In single channel wires, on the
other hand, conservation laws severely restrict the phase space
available for scattering processes, and pair collisions do not
provide a relaxation mechanism. If electron density is not too low
then three-particle collisions~\cite{Lunde,Sirenko} constitute the
leading order relaxation process. Unlike the two-particle collision
integral in higher dimensional systems, a spectral analysis of the
corresponding three-particle collision integral in single channel
wires so far is missing. This paper aims to fill this gap. We show
that for specific momentum configurations of the scattering states,
relevant for energy relaxation, the collision integral of spinless
electrons can be diagonalized analytically. We find that zero modes
of the collision integral are separated by a gap from a continuum of
decaying modes. The gap value provides us the time scale at which
counterpropagating electrons, exposed to a small temperature
differences, thermalize.

{\it Formulation of the problem.---}Our motivation is to identify
the time-scale $\tau_{\rm th}$ at which thermalization between
out-of-equilibrium counterpropagating electrons occurs in a clean
single channel quantum wires. We pursue this goal by studying the
spectrum of the three-particle collision integral under the
assumption that left- and right-moving electrons inside the wire are
initially at distinct equilibria, characterized by different
temperatures, with $\Delta T$ being the temperature difference.

Within the Boltzmann kinetic equation approach for the electron
distribution function,
$\dot{\mathcal{F}}=-\mathcal{I}\{\mathcal{F}\}$, microscopic details
of relaxation process are stored inside the collision integral
$\mathcal{I}\{\mathcal{F}\}$. In the following we specify
$\mathcal{I}\{\mathcal{F}\}$ for the spinless electrons with
quadratic energy dispersion, interacting via the Coulomb potential
$V(x)$~\cite{SM}. In the high density limit interactions are weak,
$e^2/\hbar v_F\kappa\ll 1$, here $v_F$ is the Fermi velocity and
$\kappa$ is dielectric constant of the host material, and the
leading order relaxation process is due to three-particle
collisions:
\begin{align}\label{Coll-Int}
\mathcal{I}\{\mathcal{F}&\}_{p_1} = - \sum_{p_2,p_3}
\sum_{p_{1'},p_{2'},p_{3'}}W^{1'2'3'}_{123}
\nonumber\\
&\left[\mathcal{F}_1\mathcal{F}_2\mathcal{F}_3 \mathcal{F}^h_{1'}
\mathcal{F}^h_{2'}\mathcal{F}^h_{3'} \right.
- \left.\mathcal{F}_{1'}\mathcal{F}_{2'}\mathcal{F}_{3'}
\mathcal{F}^h_{1}\mathcal{F}^h_{2}\mathcal{F}^h_{3}\right]\,.
\end{align}
Higher order scattering processes are suppressed by the small
interaction strength and phase space. In Eq.~\eqref{Coll-Int} we
introduced notations, $\mathcal{F}_{i}=\mathcal{F}(t,x,p_i)$, and,
$\mathcal{F}^h_{i}=1-\mathcal{F}_{i}$. The
$W^{1'2'3'}_{123}=\frac{2\pi}{\hbar}|\langle
1'2'3'|\hat{V}\hat{G}_0\hat{V}|123\rangle_c|^2\delta(E_i-E_f)$ is
the rate of three-particle scattering from the incoming states
$p_{1,2,3}$ into the outgoing states $p_{1',2',3'}$ with energies
$E_{i(f)}=\sum^3_i \varepsilon_{i(i')}$, respectively. An explicit
form of $W$ for a generic two-body interaction potential
$\hat{V}=\frac{1}{2L}\sum_{k_1k_2q}V_q\hat{c}^\dag_{k_1+q}\hat{c}^\dag_{k_2-q}
\hat{c}_{k_2}\hat{c}_{k_1}$, where $\hat{c}^\dag_k$ ($\hat{c}_k$) is
the electron creation (annihilation) operator, has been recently
derived in Ref.~[\onlinecite{Lunde}]. We here merely mention that
$\hat{G}_0$ denotes the free particle Green's function, subscript
``c'' refers to irreducible scattering processes, $L$ is wire length
and $V_q$ is Fourier transform of interaction potential $V(x)$.

Because of nonlinearity of the collision integral in
Eq.~\eqref{Coll-Int} an exact analytical solution of the Boltzmann
equation is very difficult to find. A simplification is possible
within the linear response analysis in which distributions entering
Eq.~\eqref{Coll-Int} can be linearized around an equilibrium state,
$\mathcal{F}_p\!=\!f_p+f_p f_p^h\psi_p$. Here
$f_p\!=[e^{(\varepsilon_p-\varepsilon_F)/T}+1]^{-1}$ is equilibrium
Fermi distribution with $\varepsilon_F$ the Fermi energy and
temperature $T$, $f_p^h\!=\!1-f_p$, and $\psi_p$ in response to
externally applied perturbation, in our case $\psi_p\! \propto\!
(\varepsilon_p\!-\!\varepsilon_F){\rm sgn}(p)\Delta T$. Restricting
to the linear response regime we insert above distribution into the
collision integral Eq.~\eqref{Coll-Int} and arrive at the Boltzmann
equation, $\dot{\psi}_{p_1}\!=-{\cal L}\{\psi\}_{p_1}$, with the
linear collision operator
\begin{equation}\label{Coll-Int-Linear}
{\cal L}\{\psi\}_{p_1} =\frac{1}{f_1f^h_1}\sum_{p_2,p_3\atop
p_{1'},p_{2'},p_{3'}} \mathcal{K}_{\{p_i\};\{p_{i'}\}}\sum^{3}_{i=1}
(\psi_{p_i}-\psi_{p_{i'}})\,.
\end{equation}
Here the kernel
$\mathcal{K}_{\{p_i\};\{p_{i'}\}}=W^{1'2'3'}_{123}f_1f_2f_3f^h_{1'}f^h_{2'}f^h_{3'}$.
A spectral analysis of the linearized collision operator in
Eq.~\eqref{Coll-Int-Linear} under the above formulated assumptions
is the problem we address in the following.

{\it Zero modes and symmetries.---}Eigenfunctions of  the collision
integral with eigenvalue zero (``zero modes'') correspond to
constant in time solutions of the Boltzmann equation and are
associated with the conserved quantities in the system. Indeed, it
is readily checked that ${\cal L}$ in Eq.~\eqref{Coll-Int-Linear} is
nullified by $\psi_{E}=\varepsilon_p$ (energy conservation),
$\psi_P=p$ (momentum conservation), and $\psi_N={\rm const}$
(conservation of total particle number). Since we are interested in
thermalization we can further restrict to processes in which all of
the participating states are close to the Fermi points. Then the
difference in number of left- and right-moving electrons, $\Delta
N$, is also conserved and $\psi_{\Delta N}=\mathrm{sign}(p)$ is an
additional zero mode. We only briefly mention that three-particle
collisions changing $\Delta N$ require backscattering and are
important for the relaxation of differences in the chemical
potentials of counterpropagating electrons~\cite{JTKA}.

Defining the Hilbert space of functions endowed with scalar product
$\langle\psi_p|\psi'_p\rangle=\frac{1}{2mT}\int^{+\infty}_{-\infty}dp\,
f_pf^h_p\psi_p\psi'_p$ it is readily checked that ${\cal L}$ is
positive Hermitian, implying a spectrum of eigenvalues larger or
equal to zero. The zero modes form a basis of the four-dimensional
subspace of conserved quantities. Any $\psi_p$ that falls off this
category evolves according to the Boltzmann equation and eventually
relaxes into one of the zero modes or their linear combination. In
general, the collision operator ${\cal L}$ may have discrete and
continuous parts of the spectrum. However, only if the zero-modes
are separated by a well-defined gap to a smallest nonvanishing
eigenvalue the concept of a relaxation rate is justified.

It is helpful to account for an additional symmetry of the
linearized collision integral. As a consequence of invariance of the
scattering kernel under reversal of momenta,
$\mathcal{K}_{\{p_i\};\{p_{i'}\}}=
\mathcal{K}_{\{-p_i\};\{-p_{i'}\}}$, $\mathcal{L}$ commutes with the
inversion operator  $\Pi \psi_p=\psi_{-p}$. Therefore,
eigenfunctions of ${\cal L}$ have well-defined parity and the
operator itself can be decomposed into a direct sum ${\cal L}={\cal
L}^+\oplus {\cal L}^-$ of operators ${\cal L}^\pm$ acting in the
mutually orthogonal subspaces of even and odd-parity functions.
(Notice that the Hilbert space defined above decomposes into a
direct sum of even and odd-parity functions, mutually orthogonal to
each other.) As we are interested in the relaxation of an odd-parity
perturbation, $\psi_{-p}=-\psi_p$, our main focus is on the spectrum
of ${\cal L}^-$.

\begin{figure}
  \includegraphics[width=8cm]{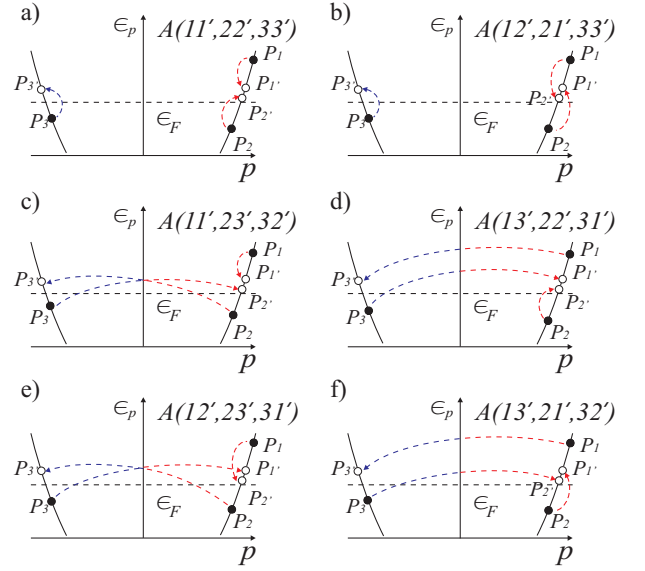}\\
  \caption{[Color online] Three-particle scattering processes that
  allow for the energy exchange between counterpropagating
  electrons and thus lead to their thermalization. The
 matrix elements $\langle1'2'3'|\hat{V}\hat{G}\hat{V}|123\rangle_c$
  decompose into six contributions, $A(11',22',33')$ plus five
  permutations of primed arguments, corresponding to the direct $a)$ and
  five exchange terms $b)$--$f)$.}
  \label{Fig1}
\end{figure}

{\it Spectrum of the linearized collision integral.---}We start
analysis from identifying small parameters in the problem. Fermi
blocking in combination with conservation laws restricts
participating scattering states to momentum strips of order $\delta
p\sim T/v_F\ll p_F$ around the Fermi points. Denoting
$q_i=p_{i'}-p_i$ the momentum transfer in a collision we thus have
the small parameters $|q_i/p_j|\sim T/\varepsilon_F\ll1$, where
$i,j=1,2,3$. The energy is transferred via the three-particle
collision in which one of the incoming electrons, say with momentum
$p_3$, scatters off two other counterpropagating electrons (see
Fig.~\ref{Fig1} for the six possible scattering processes where an
electron with $p_3<0$ scatters off two right movers). Using momentum
conservation $q_2+q_1+q_3=0$ we can express energy conservation as
$p_2(q_1 + q_3) - (p_1 q_1 + q_1^2 + p_3 q_3 + q_1 q_3 + q_3^2)=0$
which, solved for $q_3$, shows that $q_3=q_1(p_1-p_2)/2p_F + {\cal
O}\left((p_1-p_2)^2/p^2_F,q^2_1/p_F^2\right)$. It becomes clear now
that energy transfer between the counterpropagating electrons occurs
via small portions of momentum exchange $v_F|q_3|\sim
T^2/\varepsilon_F$, and $\{|q_3/q_1|, |q_3/q_2|\}\ll 1$ present
additional small parameters, while $q_1\simeq -q_2$ up to
corrections of ${\cal O}\left(q_2(p_1-p_2)/p_F)\right)$.

A further important ingredient for our calculation is the
three-particle scattering rate for the Coulomb potential. Expanding
$W^{1'2'3'}_{123}
=\Lambda^{1'2'3'}_{123}\delta(E_i-E_f)\delta_{q_1+q_2+q_3=0}$ to the
leading order in the above small parameters, we find that the
scattering rate depends only logarithmically on the momenta transfer
$q_i$. Indeed, $\Lambda^{1'2'3'}_{123} ={2\pi\over \hbar}({2e^2\over
\kappa})^4 [ {(k_Fw)^2 \over 2L^2\varepsilon_F} ]^2
\Gamma^2(q_1,q_3)$, with
\begin{align}
\label{gamma} \Gamma(q_1,q_3)= \left[ 1 + 3 (\gamma_E + \ln [k_F w]
)\right] \ln[ 2p_F|q_3|/ q_1^2].
\end{align}
where $\gamma_E$ is the Euler constant. This result
[Eq.~\eqref{gamma}] applies at not too low temperatures $\{T d/\hbar
v_F, k_Fd\}\gg1$ at which screening of the nearby gate is not
important. Technically speaking, the typical Fourier component $V_k$
of the interaction potential then has wave number in the range
$d^{-1}\ll |k| \ll w^{-1}$, so that $V_k\simeq {2e^2\over \kappa}
\left[ \ln{2e^{-\gamma_E}\over |k|w} + {(kw)^2\over 4}
\ln{2e^{1-\gamma_E}\over |k|w}\right]$~\cite{SM}. The logarithmic
$q_i$-dependence of the scattering rate is a result of subtle
cancellations between direct and various exchange terms contributing
to the scattering amplitude, see also Fig.~\ref{Fig1}. It is owing
to this property of  $W$ that makes an analytical diagonalization of
${\cal L}$ possible.

Our strategy is now to split the linearized collision operator
into two contributions
\begin{equation}
{\cal L}\{\psi\} = {\cal L}_0\{\psi\}+ \delta {\cal L}\{\psi\}\,,
\end{equation}
where ${\cal L}_0$ allows for an analytical diagonalization, while
corrections to the spectrum from $\delta {\cal L}$ turn out to be
irrelevant for our problem (and may, in principle,  be calculated
from perturbation theory). More precisely, we choose ${\cal L}_0$ as
in Eq.~\eqref{Coll-Int-Linear} with kernel $\mathcal{K}_0$ resulting
from the original $\mathcal{K}$ upon linearization of  the quadratic
energy dispersion, i.e. upon substituting $f_{p\pm p_F}$ and
$\delta(E_i-E_f)$, respectively, by $g_{\pm p}\equiv(e^{\pm v_F
p/T}+1)^{-1}$, and $\delta_0(E_i-E_f)\equiv {L\over 2hv_F} \left[
\delta_{q_1+q_2} \delta_{q_3}  \Theta(p_1,p_2,-p_3) +
\delta_{q_2+q_3}\delta_{q_1} \Theta(p_1,-p_2,-p_3) \right]$,
$\Theta(p_1,p_2,-p_3) =\left[ \theta(p_1)\theta(p_2)\theta(-p_3) +
\theta(-p_1)\theta(-p_2)\theta(p_3) \right]$. That is the operator
${\cal L}_0$ corresponds to a problem with linear dispersion
relation. The expectation that corrections to the spectrum from
$\delta\mathcal{L}$ are small is, of course, a consequence of the
fact that three-particle scattering processes which provide energy
exchange involve all colliding electrons near the Fermi points,
while states at the band bottom are not crucial for thermalization.
Since the above approximation preserves inversion symmetry, $[{\cal
L}_0,\Pi]=0$, eigenfunctions of ${\cal L}_0$ have well-defined
parity. Given this property we may restrict the eigenvalue equation
${\cal L}_0\{\psi^n\}_{p}=\omega_n \psi^n_{p}$ to momenta $p>0$ and
then extend solutions to negative values $p<0$ by taking even and
odd parity combinations
$\psi_{p}=\theta(p)\psi^n_{p}\pm\theta(-p)\psi^n_{-p}$.

We next show that ${\cal L}_0$ can be transformed into a linear
second-order differential equation. As a first step, we adopt
logarithmic accuracy approximation to substitute the argument of the
logarithm in the momentum dependent scattering rate
Eq.~\eqref{gamma} by its typical value,
$q_1^2/(q_3p_F)\rightarrow1$, as dictated by the conservation laws.
The linear collision operator with constant scattering rate can be
reduced then to
\begin{eqnarray}
&&\hskip-.5cm \label{L1} {\cal L}_0\{ \psi \}_{p_F+p_1}= \gamma_0
\Big[ \frac{L}{2h} \left( \varkappa^2+p^2_1\right)
\psi_{p_F+p_1}  \nonumber \\
&&\hskip-.5cm- \frac{1}{g^h_{p_1}}  \sum_{p_2}  (p_2-p_1)
g^h_{p_2}b_{p_2-p_1} \left( 2 \psi_{p_F+p_2}-\psi_{p_F-p_2} \right)
\Big] ,
\end{eqnarray}
where $\varkappa=\pi T/v_F$ is characteristic momentum due to
thermal smearing of Fermi functions, $b_p=(e^{v_Fp/T}-1)^{-1}$ is
the bosonic distribution function, $g^h_p=1-g_p$, and $\gamma_0\!=\!
{(Lk_F)^2\over h^2}{T L\ln^22\over \varepsilon^2_F} ({2e^2\over
\kappa})^4 [{(k_Fw)^2 \over 2L^2\varepsilon_F }]^2 \ln^2[k_F w]$.
The sequence of exact transformations leading to Eq.~\eqref{L1} can
be summarized as follows~\cite{SM}. (i) It is convenient to organize
${\cal L}_0$ into six contributions ${\cal
L}_0=\sum_{s=\pm}\sum_{j=1}^3 l_j^{s-}$ where
$l_j^{s-}\{\psi\}=\frac{1}{g_1g^h_1}\sum_{p_2p_3}\sum_{p_{1'}p_{2'}p_{3'}}
\mathcal{K}^{s-}_0 [\psi_j-\psi_{j'}]$ and $\mathcal{K}_0^{+-}$
($\mathcal{K}_0^{--}$) describes processes in which the right-mover
$p_1$ scatters off one right- and one left-mover (two left-movers).
(ii) In the individual contributions  $l_j^{s-}$ we remove two out
of five momentum-integrations employing conservation laws and
complete two further integrations with help of the identities:
$\sum_pg_pg^h_{p-q}= \frac{L}{h} q b_q$, $\sum_q
g_{p+q}g^h_{p-q}=-\frac{L}{h} p b_{-p}$, and $\sum_q q
g_{p+q}b_q=-\frac{L}{h} (\varkappa^2+p^2)g_p$. (iii) We find  that
all terms $l_j^{--}\{\psi\}$ and $l_3^{+-}\{\psi\}$ are identical
zero, while
 $l^{+-}_{1}\{\psi\}+l^{+-}_{2}\{\psi\}$ can be summed to
give Eq.~\eqref{L1}.

We then introduce odd and even momentum combinations
with respect to the Fermi point
\begin{equation}
\psi^{\pm}_{p}=\sqrt{g_p g^h_p}
\big[\psi_{p_F+p} \pm \psi_{p_F-p}\big]\,,
\end{equation}
where the normalization factor is chosen for convenience, and recast
the eigenvalue problem for odd and even combinations, $s=\pm$, as
follows
\begin{equation}
\label{psi-even/odd} \omega_n \psi^{ns}_{p_1}\!\! = \! \gamma_0
\left[ \frac{L}{2h}(p^2_1 + \varkappa^2) \psi^{ns}_{p_1}\! -
\!\!\sum_{p_2} \frac{ 3^{\delta_s} (p_2-p_1)
\psi^{ns}_{p_2}}{2\sinh\frac{v_F(p_2-p_1)}{2T}}\right],
\end{equation}
with $\delta_\pm=0,1$, respectively. At this stage we introduce
dimensionless momentum $k=\frac{v_F p}{\pi T}$, energy
$\Omega_n={2h\over \gamma_0 L } \left({\pi v_F\over T}\right)^2
\omega_n$, and notice that the kernel in Eq.~\eqref{psi-even/odd}
depends on the difference of its arguments, which makes it
convenient to perform a Fourier transformation
$\psi^{n\pm}_{k}=\int\frac{dx}{2\pi}\psi^{n\pm}_{x}e^{ikx}$ and
$\int \frac{ke^{ikx}dk}{\sinh(\pi k/2)}=\frac{2}{\cosh^2x}$. As a
result, the eigenvalue equations for the Fourier images
$\psi^{ns}_k$ reduce to Schr\"odinger equations of a particle moving
in a P\"oschl-Teller potential
\begin{eqnarray}
\label{pt} &&\left[ \frac{d^2}{dx^2}+\Omega_n-1+\frac{2\cdot
3^{\delta_s} }{\cosh^2x}\right] \psi^{ns}_{x}=0,
\end{eqnarray} which
can be solved with help of operator-algebra technique known
from the harmonic oscillator problem~\cite{Schwabl}.

The eigenvalue problem in Eq.~\eqref{pt} allows for one even- and
one odd-parity bound-state $\Omega_n=0$ of the form
$\psi^{0+}_{x}=1/\cosh x$ and $\psi^{0-}_{x}=-3\sinh x/\cosh^2x$,
respectively~\cite{fn1}. Upon inverse Fourier transformation and
extension to negative momenta in the above prescribed manner this
gives the four zero-modes $\psi^{0+}_{N}={\rm const}$,
$\psi^{0-}_{\Delta N}=\mathrm{sgn}(p)$, $\psi^{0-}_P=p$, and
$\psi^{0+}_{E}=|p|$. As already discussed, the first three functions
are consequences of conservation of total particle number, momentum
and the difference in number of left- and right-moving electrons.
The fourth zero mode expresses conservation of energy for the
linearized spectrum. Of course, these four zero modes could have
been directly inferred from ${\cal L}_0$.

More relevant for our problem is the fact that these bound-state
solutions are separated by a gap $\delta\Omega=1$ from a continuum
of propagating modes and ${\cal L}_0$, therefore, possess a
well-defined smallest nonvanishing eigenvalue. In order to associate
this latter with the thermalization rate, we have to make sure that
this gap is also present in the original collision operator ${\cal
L}^-$. Employing that eigenfunctions of ${\cal L}_0$ form a complete
set we may express ${\cal L}^-$ in terms of its odd-parity subset.
Finally, reminding that $\psi_{\Delta N}$ and $\psi_P$ are also
nullified by ${\cal L}^-$, it is evident that the smallest
nonvanishing eigenvalue of ${\cal L}^-$ is of order $\delta\omega =
{\rm min}_{\psi^-_{k}}\{ \langle\psi^-_{k} | {\cal L} |\psi^-_{k'}
\rangle \}$, with $\{\psi^-_{k}\}$ the set of eigenstates
corresponding to odd-parity propagating solutions of Eq.~\eqref{pt}.
Since matrix elements $|\langle\psi^-_{k} | \delta {\cal L}
|\psi^-_{k'} \rangle| \ll |\langle\psi^-_{k} | {\cal L}_0
|\psi^-_{k'} \rangle|$ it readily follows that ${\cal L}^-$ and
${\cal L}_0^-$ share a gap of same order. Employing then $\delta
\Omega =1$, restoring original units,  and inserting the explicit
form of $\gamma_0$ we arrive at the thermalization rate
\begin{equation}
\label{tth} 1/\tau_{\mathrm{th}}=c(\varepsilon_F/\hbar)(e^2/\hbar
v_F\kappa)^4\lambda^2(k_Fw)(T/\varepsilon_F)^3\,,
\end{equation}
where coefficient $c=\ln^22/2\pi^6$ and function $\lambda(x)=x^2\ln
x$. Equation~\eqref{tth} presents the main result of this paper. We
note here that the temperature dependence of $\tau_{\rm th}^{-1}$
can be understood from a simple phase-space argument like in Fermi
liquid theory. To first approximation the thermalization rate
follows from the out-scattering part of the collision integral,
$1/\tau_{\rm out} \propto \sum_{\{p_i\}}\mathcal{K}_{\{p_i\}}$,
where two out of the five momenta $\{p_i\}$ are fixed by the
conservation laws, while the remaining three extend over momentum
range set by the temperature broadening $\delta p\sim T/v_F$ of
Fermi distributions. Since the scattering amplitude is only
logarithmically dependent on momentum transfer then scattering rate
$W \simeq {\rm const}$ and, therefore, $1/\tau_{\rm out} \propto
(T/\varepsilon_F)^3 $~\cite{fn2}.

{\it Conclusions.---} We have analyzed the spectrum of the
three-particle collision integral in a one-dimensional electron
system. We found that zero modes, associated with the conservation
laws, are separated by an energy gap from a continuum of propagating
modes, and identified the gap with the relaxation rate $\tau_{\rm
th}$ relevant for thermalization of counterpropagating electrons.
Our analysis applies to clean single channel quantum wires of
spinless electrons at not too low densities. It is highly desirable,
yet very challenging, to extend the analysis to lower densities
where interactions become strong~\cite{MAP}. Also, inclusion of spin
\cite{torsten} presents an interesting problem, since at very low
temperatures the effect of spin-charge coupling becomes
relevant~\cite{SCS}.

{\it Acknowledgements.---} We are sincerely grateful to
K.~A.~Matveev for discussions which stimulated this project. We
would like to acknowledge also useful communication with P.~Brouwer,
A.~Imambekov, L.~Glazman, A.~Kamenev, T.~Karzig, F.~von~Oppen and
Z.~Ristivojevic. The work at Argonne National Laboratory was
supported by the U.S. DOE, Office of Science, under Contract No.
DEAC02-06CH11357.


\begin{thebibliography}{99}
\bibitem{fn1}
Notice that the eigenvalue equation for $s=-$ allows for the second
bound-state solution $\psi^{-,2}(x)- = 3/ \cosh^2x$  with energy
$\lambda_n=-3$. This is, however, an even parity function.
\end{thebibliography}
\end{document}